\documentclass{aa}

\usepackage{rotate} 
\input rotate.tex

\def\deg{\hbox{$^\circ$}}

\def\fdg{\hbox{$.\!\!^\circ$}}

\def \hi {H~{\sc i~}} 
\def \hii {H~{\sc ii~}} 
\def\kms{km\,s$^{-1}$}

\begin{document} 
\thesaurus{10(09.03.1; 10.11.1; 11.12.1)} 
\title{The Kinematic and Spatial Deployment of Compact, Isolated
  High-Velocity Clouds} 
\titlerunning{Compact, Isolated
  High-Velocity Clouds} 
\author{R. Braun\inst{1} and W. B. Burton\inst{2} } 
\institute{Netherlands Foundation for Research in Astronomy, P.O. Box 2, 
7990 AA Dwingeloo, The Netherlands 
\and Sterrewacht Leiden, P.O. Box 9513, 2300 RA Leiden, The Netherlands } 

\date{Received mmdd, 1998; accepted mmdd, 1998} 
\offprints{R. Braun or W.B. Burton} 
\maketitle 

\begin{abstract} 
  
  We have identified a class of high--velocity clouds which are
  compact and apparently isolated.  The clouds are compact in that
  they have angular sizes less than 2 degrees {\sc FWHM}. They are isolated
  in that they are separated from neighboring emission by expanses
  where no emission is seen to the detection limit of the available
  data.  Candidates for inclusion in this class were extracted from
  the Leiden/Dwingeloo \hi survey of Hartmann \& Burton
  (\cite{hart97}) and from the Wakker \& van Woerden (\cite{wakk91})
  catalogue of high--velocity clouds identified in the surveys of
  Hulsbosch \& Wakker (\cite{huls88}) and of Bajaja et al.
  (\cite{baja85}).  The candidates were subject to independent
  confirmation using either the 25--meter telescope in Dwingeloo or
  the 140--foot telescope in Green Bank.  We argue that the resulting
  list, even if incomplete, is sufficiently representative of the
  ensemble of compact, isolated HVCs -- CHVCs -- that the
  characteristics of their disposition on the sky, and of their
  kinematics, are revealing of some physical aspects of the class.
  The sample is more likely to be representative of a single
  phenomenon than would a sample which included the major HVC
  complexes.  We consider the deployment of the ensemble of CHVCs in
  terms used by others to ascertain membership in the Local Group, and
  show that the positional and kinematic characteristics of the
  compact HVCs are similar in many regards to those of the Local Group
  galaxies.  The velocity dispersion of the ensemble is minimized in a
  reference frame consistent with the Local Group Standard of Rest.
  The CHVCs have a mean infall velocity of 100~km~s$^{-1}$ in the
  Local Group reference frame. These properties are strongly
  suggestive of a population which has as yet had little interaction
  with the more massive Local Group members. At a typical distance of
  about 1~Mpc these objects would have sizes of about 15~kpc and gas
  masses, M$_{\rm HI}$, of a few times 10$^7$~M$_\odot$, corresponding to
  those of (sub-)dwarf galaxies.

\end{abstract} 
\section{Introduction} 
\label{intro}

The term High Velocity Cloud (HVC) has traditionally been assigned to
features detected in the $\lambda$21~cm line of neutral hydrogen
emission which (1) have a radial velocity which is discrepant from
that due to Galactic rotation in the direction of the feature (Muller
et al. \cite{mull63}) and (2) subsequently prove to have no obvious stellar
counterpart. Aberrant velocity \hi features which do have a detected
stellar counterpart are, of course, classified as galaxies.  It is
obvious that this empirical definition of an HVC is not
altogether satisfactory, but it underlines the fact that despite some
35 years of study, no generally accepted explanation for the origin,
the distance and hence the basic properties of these enigmatic
features has yet emerged.  The class is effectively defined by {\it
what they are not}. These are features which are {\it not} related
in a simple way to the bulk of Galactic \hi via spatial or kinematic
continuity and for which we are {\it not yet aware\ } of an associated
stellar or gaseous component. This last point in particular is one
that has by no means been addressed exhaustively and may well prove to
be a fruitful avenue for future study. Indeed, recent efforts to
detect associated diffuse ionized gas in H$\alpha$ emission are
beginning to meet with some success (Reynolds et al.  \cite{reyn98}).

It must be stressed that the term HVC encompasses a
fairly wide range of observational phenomena. The distinct system of
\hi features known as the Magellanic Stream was first recognized by
Mathewson, Cleary \& Murray (\cite{math74}) to represent tidal debris
originating in the gravitational interaction of the Galaxy with both
the Large and Small Magellanic Clouds. This system subtends an
extremely large solid angle, defining a belt that encircles the entire
Galaxy. Given the spatial deployment and kinematic continuity of this
system it can be plausibly assigned a distance of some tens of kpc.
Some of the well-studied HVC complexes (known as A, C, H and M) are
also distributed over large regions of some tens of square
degrees. The distances of two of these complexes have been constrained
by absorption--line observations: Danly et al.  (\cite{danl93})
determined Complex M to lie within the distance range $1.7 < d < 5$
kpc, and van Woerden et al. (\cite{woer98}) found Complex A to lie
within the range $4 < d < 10$ kpc.  These absorption--line
observations, toward isolated lines of sight in two HVC complexes, are
the only direct measures of HVC distances available. At the other
extreme, is a class of object that appears to be intrinsically
compact, with angular dimensions of a few degrees at most. These
objects have no obvious relation either spatially or in velocity with
the extended complexes and include the most extreme LSR
velocities. 

It is this last category of compact, peculiar velocity \hi emission
features which we consider here. The lack of obvious
association with either the Galaxy or the large HVC complexes and
their similar angular size may allow definition of a single class of
HVCs, whose members plausibly originated under common circumstances
and share a common subsequent evolutionary history.  By utilizing the
Leiden/Dwingeloo Survey of \hi emission of Hartmann \& Burton
(\cite{hart97}) and the HVC catalog of Wakker \& van Woerden
(\cite{wakk91}) we have compiled a list of candidate sources.
Confirming observations were obtained of all candidates. A
representative sample of 66 compact, isolated HVCs (or CHVCs) has
emerged, including 23 sources cataloged here for the first time. The
sample is distributed quite uniformly over the sky, and yet defines a
well-organized kinematic system. The kinematic signature of this
system suggests a strongly in-falling population associated with the
Local Group gravitational potential. At a typical distance of 1~Mpc
they have the dimensions (15~kpc) and gas masses
(M$_{\rm HI}~\sim~10^{7.5}$M$_\odot$) of (sub-)dwarf galaxies. A similar
suggestion was made by Blitz et al. (\cite{blit98}) based on a more
heterogeneous sample of HVCs.

Our discussion is organized as follows.  We begin by describing the
method of sample selection in \S\ref{sec:samp}, proceed with a
description of the newly acquired observations in \S\ref{sec:data} and
continue with a presentation and discussion of our results in
\S\S\ref{sec:results} and \ref{sec:disc}.

\section{Sample Selection} 
\label{sec:samp}

The Leiden/Dwingeloo Survey (hereafter abbreviated as the LDS,
Hartmann and Burton \cite{hart97}) of \hi emission in the Northern sky
above declination about $-30^\circ$ has made it possible to select a
well-defined sample of candidate sources of small angular size. This
survey provides nearly uniform spatial sampling on a 0\fdg5 grid
at a sensitivity of about T$_{\rm B}$~=~0.07~K rms at a velocity
resolution of 1~km~s$^{-1}$ over the V$_{\rm LSR}$ range of $-450$ to
$+400$~km~s$^{-1}$. When smoothed to the typical HVC velocity width of
20~km~s$^{-1}$, the LDS sensitivity is about 0.02~K rms. The earlier
HVC survey of Hulsbosch \& Wakker (\cite{huls88}) already provided
spatial sampling on a 1$^\circ$ grid above $-17^\circ$ declination
with an rms sensitivity of about T$_{\rm B}$~=~0.01~K rms at a
velocity resolution of 16.5~km~s$^{-1}$ over the V$_{\rm LSR}$ range
of $-950$ to $+800$~km~s$^{-1}$. The southern sky coverage of
Hulsbosch \& Wakker was supplemented with data taken on a 2\deg
grid at about half the sensitivity by Bajaja et al.  \cite{baja85}. A
catalog based on these two HVC databases was presented by Wakker \&
van Woerden (\cite{wakk91}, hereafter WW91).  Since both of the
northern surveys were carried out with the Dwingeloo 25~m telescope,
with a beam size of about 35 arcmin, neither is Nyquist sampled, but
no sources should slip completely through the cracks of the LDS
sampling.

An issue of particular concern in a single-coverage survey with a
total power instrument, is that of the effect of intermittent radio
frequency interference (RFI) on the results. As noted by Hartmann
(\cite{hart94}), RFI is often of the form of extremely narrow-band
signals which are easily recognized as artificial when seen in a
moderately high-resolution spectrum. However, some types of
intermittent emission of unknown origin are characterized by a broad
spectral signature which is impossible to distinguish from a naturally
occurring spectral line profile. Only repeated total-power or
interferometric observations can be used to distinguish such broad
events from naturally ocurring ones.

We have searched the LDS database for all compact, isolated \hi
emission features. The method used to identify candidates was to
construct a series of ``channel maps'' of \hi emission, each
integrated over a velocity interval of 32~km~s$^{-1}$ and spanning the
entire spatial coverage of the survey ($\delta>-30^\circ$). Compact
features (less than 2$^\circ$ {\sc FWHM}) were cataloged which were
spatially and kinematically distinct from Galactic \hi emission.  The
condition of kinematic distinction from the Galaxy corresponds
approximately to deviation velocities, as defined by Wakker
(\cite{wak91}), of about 50~km~s$^{-1}$ or greater.  A total of 123
candidates was then subject to new, confirming observation as
described in \S\ref{sec:data}, either using the Dwingeloo 25--meter
telescope or, for the weaker of the candidates, the 140--foot
telescope of the NRAO in Green Bank.

The 44 candidates whose reality was confirmed in the new data entered
the tabulation of CHVCs. Some of the candidates on the initial list
extracted from the Leiden/Dwingeloo survey could not be confirmed. In
some cases, the initial candidate was 
only of marginal signal--to--noise ratio. In other cases the
second--epoch Dwingeloo spectra were not of sufficient quality to
confirm the feature. We note that the interference environment in
Dwingeloo had deteriorated substantially during the period between
1993, when the last observations for the LDS were made, and late 1997
and early 1998 when the confirmations were attempted.  However, a
number of the candidate sources turned out to
correspond to cases of non-repeatable broad-band interference. Many of
the sources which we could confirm were subsequently imaged with small
Nyquist-sampled pointing grids to more accurately determine source
positions and sizes as described below.

The first step in compiling the list of compact, isolated HVCs was
based on inspection of the Leiden/Dwingeloo survey alone, without
consulting any earlier material.  In a second step, we consulted the
catalog of Wakker \& van Woerden (\cite{wakk91}), extracting from it
all entries identified with a value of the parameter $N \leq 3$, i.e.
referring to HVC detections based on an observation at a single
pointing, or, at most, on 3 spectra and therefore having a surface
area of less than about 4 square degrees.  This is true of fully 430
of the total of 561 HVCs cataloged in WW91. We extracted LDS spectra
at each of these 430 HVC positions. Of these, 59 had no data in the
LDS, since they were below the $-30^\circ$ southern declination limit,
94 could not be confirmed to the sensitivity of the LDS, 62 had a
marginal confirmation and 216 could be confirmed with some confidence.
Subsequently, we produced images of integrated HI emission at the
position of each of the 216 confirmed WW91 sources integrated over a
velocity range of 100~km~s$^{-1}$ centered on the cataloged velocity
centroid. Although all of these sources represent real \hi emission
features at a peculiar velocity, most of them appear to represent
substructure of more extensive HVC complexes. We have extracted only
those sources which could be classified as both compact and isolated
by demanding: (1) that the half power contour of each source in our
images of integrated \hi emission had an area of less than 4 square
degrees and (2) that to the sensitivity limit of the LDS the source was
not connected to a diffuse emission complex. Only 42 of the 216
confirmed WW91 sources could satisfy those criteria and only 22 of
these had not already been included in our sample.

In a final step, we consulted other publications which had reported
HVCs which met our criteria of compactness and isolation, but found
that these few sources, referenced in column 13 of the table, had
already been recovered during the first two steps.  We note in this
regard that Davies's (\cite{davi75}) cloud, {\sc hvc}\,120$-$20$-$444,
was excluded by our isolation criterion -- it is less than $1^\circ$
removed from intense emission from M31 -- whereas we have no reason to
think that this cloud does not show the same intrinsic properties
characteristic of the CHVC objects in our tabulation.

\section{Data} 
\label{sec:data}

Observations were obtained with the Dwingeloo 25~m telescope in the
periods 1997 September 17 to 19, 1997 November 27 to December 15 and
1998 March 2 to 3, for a total of about 21 days. A variety of
switching methods, central frequencies, and total bandwidths was
employed in an attempt to optimize baseline quality and avoid locally
generated narrow band interference. The most successful strategy
employed position switching to a nearby reference position and a total
bandwidth of 10~MHz centered within a few hundred km~s$^{-1}$ of the
Local Standard of Rest. The correlator provided 1024 spectral channels
across the band. The single-polarization receiver had a typical system
temperature of 35~K.  Spectra were calibrated in amplitude with
regular observations of the standard region S8, at
($l,b$)=(207\fdg0,$-$15\fdg0), for which a peak line brightness of
T$_B$~=~71~K and line integral of 840~K~km~s$^{-1}$ over the V$_{LSR}$
interval $-5$ to $+22$ \kms was assumed (Williams \cite{will73}, Hartmann
\cite{hart94}).

Candidate source positions were initially observed with a single pair
of on-source and off-source pointings separated by two degrees in
galactic latitude. The typical on-source integration time was 6
minutes, providing an rms sensitivity of about 0.04~K at a velocity
resolution of 2~km~s$^{-1}$ in the calibrated spectra. As time
permitted, confirmed sources were reobserved with a $3\times3$ grid
with a pointing separation of 15 arcmin and subsequently with an
additional $3\times3$ grid with a 30 arcmin pointing separation.

Additional observations were made with the Green Bank 140--foot
telescope of the NRAO in 1997 September and 1997 December. A total
bandwidth of 10~MHz was employed centered near the candidate
frequency. The correlator provided 512 spectral channels in each of
two polarizations. The typical system temperature was
20~K. Calibration spectra were obtained on the standard region
S8. The observations were made in frequency switched mode with a
10~MHz switch to higher frequency. Typical on-source integration times
were less than about 5 minutes.

\section{Results}
\label{sec:results}

Our catalog of 66 isolated CHVCs is shown in Table~\ref{tab:T_data}. The
columns of the table are defined in the following way.

Column~1: Running identifying number in the catalog.

Column~2: The integer rounded Galactic longitude and latitude together
with the integer rounded LSR velocity. All three coordinates in the
designation are derived from the best available data for each
source as outlined below. 

Column 3: I/C = Initial/Confirmation data. The initial source of the
candidate positions and velocities was either the LDS, indicated by
``LD'' in this column, or Wakker \& van Woerden (\cite{wakk91}),
indicated as ``W''. Confirming data were required in all cases to
establish the repeatability of the source spectrum. In some cases the
LDS provided independent confirmation of the ``W'' sources, while in
other cases new data were obtained with the NRAO 140-foot telescope in
Green Bank, denoted with ``GB'' or the Dwingeloo 25-m telescope,
denoted with ``D''. Those cases where Nyquist sampled maps were made
with the Dwingeloo telescope are indicated with the designation
``Dm''.

Columns 4 and 5: Galactic $(l,b)$ coordinates of the source centroid as
determined from the new Nyquist sampled images of the integrated \hi
where available (designated with a ``Dm'' in column 3) and otherwise
from the integrated \hi data of the LDS. Positional accuracy is about
5~arcmin for the ``Dm'' data and about 15~arcmin for the LDS.

Columns 6 and 7: J2000 right ascension and declination coordinates of
the source centroid. Positional accuracy is about 5~arcmin for the
``Dm'' data and about 15~arcmin for the LDS.

Columns 8 and 9: The radial velocity measured with respect to the
Local Standard of Rest and Galactic Standard of Rest.  For those
members of the ensemble which were subject to detailed mapping, the
$v_{\rm LSR}$ refers to the velocity at the centroid position; for the
other CHVCs, the velocity is that following from a Gaussian
decomposition of the representative spectrum plotted in
Fig.~\ref{fig:F_inthi}.  The galactic standard of rest is defined by $v_{\rm
  GSR} = 220 \cos(b) \sin(l) + v_{\rm LSR}$.

Columns 10, 11 and 12: The peak brightness temperature of the Gaussian
component resulting from decomposition of the representative spectrum
plotted in Fig.~\ref{fig:F_inthi}, the {\sc fwhm} velocity width, and the
integrated flux contributed by the component.

Columns 13, 14 and 15: The deconvolved major and minor axis {\sc FWHM}
dimensions and the major axis position angle (East of North) in
$(l,b)$ coordinates.  These are derived from the new Nyquist-sampled
images of the integrated \hi where available (designated with a ``Dm''
in column 3) and otherwise from the integrated \hi data of the LDS.

Column 16: References to earlier mentions of the tabulated HVCs, coded
as follows: {\sc bbwh}99, Burton, Braun, Walterbos, \& Hoopes
\cite{burt99}; {\sc cm}79, Cohen \& Mirabel \cite{cohe79}; {\sc g}81,
Giovanelli \cite{giov81}; {\sc gh}77, Giovanelli \& Haynes
\cite{giov77}; {\sc h}78, Hulsbosch \cite{huls78}; {\sc h}92, Henning
\cite{henn92}; {\sc m}81, Mirabel \cite{mira81}; {\sc mc}79, Mirabel
\& Cohen \cite{mira79}; {\sc w}\#, Wakker \& van Woerden
\cite{wakk91}; and {\sc w}r79, Wright \cite{wrig79}.

Each of the CHVCs is illustrated in Fig.~\ref{fig:F_inthi} where the
integrated \hi intensity from the vicinity of each source is
displayed, together with a representative \hi spectrum. The images
were obtained from an integration over 200~km~s$^{-1}$ velocity width
of the LDS data (Hartmann \& Burton \cite{hart97}),
centered near the mean velocity of the feature.  (In a few cases, the
relatively small deviation velocity of the feature required a
different setting of the velocity interval.) The data are used as they
were extracted from the Hartmann \& Burton (\cite{hart97}) CD--ROM,
i.e. with no additional baseline manipulation. The contours in the
images represent integrated intensities labelled in units of
K~km~s$^{-1}$, which can be converted to column depth in units of \hi
atoms cm$^{-2}$ by multiplying by $1.8 \times 10^{18}$, under the
usual assumption of negligible optical depth. The representative
spectrum refers to the direction in the $0.\!^\circ 5 \times
0.\!^\circ 5$ grid nearest to the peak of the integrated emission.
For those compact HVCs whose reality was confirmed in Green Bank (as
indicated in column 3 of the table), the spectrum displayed was
obtained on the 140--foot telescope; for all other entries, the
spectrum displayed is from the Leiden/Dwingeloo survey after a single
pass of Hanning smoothing.

\subsection{Comments on a few individual CHVCs} 

The entries in Table~\ref{tab:T_data} represent a range of profile shapes.
The {\sc fwhm} values, for example, range from 5.9 km~s$^{-1}$,
characteristic of a very narrow, cold \hi feature in the conventional
gaseous disk, to 95.4~km~s$^{-1}$, characteristic of some moderately 
massive external galaxies.  We remark here on some of the individual
members of the ensemble.  It is an important question, of course,
whether the tabulation of compact HVCs represents a single physical
phenomenon, or whether it contains interlopers from other classes of
objects.

One of the objects listed in the table has, in fact, been revealed as
an interloper.  The $19^{\rm th}$ entry, designated {\sc
  hvc}\,094$+$08$+$080, is a large, nearby, low--surface--brightness
galaxy, Cepheus~1, discovered during the course of this investigation.
Deep multicolor and spectroscopic optical follow--up observations
showed the presence of stars and \hii regions, and radio synthesis
interferometry confirmed that the galaxy has the optical properties
and \hi rotation signature of a low--surface--brightness spiral galaxy
(Burton et al. \cite{burt99}).  We have retained this object in our
tabulation, and in the various plots.  It is instructive to see how
the \hi properties of this interloper galaxy might differ from those
of the CHVCs, and how its spatial and kinematic deployment on the sky
might resemble that of the ensemble of CHVCs.  Warned by the the
presence of Cepheus~1 in our compilation, we searched the Digital Sky
Survey CD--ROM in the direction of each of the sources listed in the
table, but found a clear optical counterpart for no entry other than
{\sc hvc}\,094$+$08$+$080. We can not rule out that other entries
would reveal an optical counterpart in deeper optical data; indeed,
very deep optical searches are called for.  Even if no further large
galaxy lurks in the tabulation, the distinction between CHVCs and
dwarf galaxies with very weak star formation remains to be made.  We
view establishing the nature of such a distinction as an important
challenge.

We note a few of the entries individually:

\subsubsection{CHVCs with exceptionally broad, or exceptionally
  narrow, \hi lines} 
  
Most of the HVC flux studied in earlier investigations is contributed
from complexes, or from extended features; these commonly show
substantial kinematic gradients consistent with rotation
or shearing.  The CHVCs considered here, however, are largely subsumed
into a single beam; although the kinematic information is largely
unresolved, kinematic considerations in addition to the systemic
velocity remain relevant.  {\sc hvc}\,115$+$13$-$275 has a {\sc fwhm}
of 95.4~km~s$^{-1}$; the exceptional width of the feature and its
location at low $b$ suggests that a deeper optical look than that
afforded by the POSS would be appropriate, as would an \hi synthesis
observation looking for kinematic structure contributed either by
rotation of a single entity or by blending of subunits, each moving
with a different velocity.  {\sc hvc}\,114$-$10$-$430 and {\sc
  hvc}\,125$+$41$-$207 have, on the other hand, \hi signatures which
are exceptionally narrow not only for HVCs, but for any \hi emission
lines.  The line width of {\sc hvc}\,125$+$41$-$207, {\sc fwhm}$=5.9
\pm 1.6$~km~s$^{-1}$, corresponding to $\sigma_{\rm v} = 2.5 \pm
0.7$~km~s$^{-1}$, is sufficiently small that it may even be used to
constrain the kinetic temperature of the gas, $T_{\rm k}<750$~K.  It
also constrains the line--of--sight component of any rotation or shear
in a single object, as well as the range of kinematics if the object
should be an unresolved collection of subunits.
  
\subsubsection{CHVCs near the galactic equator} 

HVCs located on lines of sight traversing the gaseous disk of the
Milky Way display the horizontal component of their space motion.
Large horizontal motions are difficult to account for in terms of a
galactic fountain model (Shapiro \& Field \cite{shap76}, Bregman
\cite{breg80}).  Burton (\cite{burt97}) has noted that HVCs do not
contaminate the \hi terminal--velocity locus in ways which would be
expected if they pervaded the galactic disk, and that this observation
constrains HVCs either to be an uncommon component of the Milky Way
disk, confined to the immediate vicinity of the Sun, or else to be
typically at large distances beyond the Milky Way disk.  In our
tabulation, {\sc hvc}\,024$-$02$-$285, {\sc hvc}\,040$+$01$-$282, and
{\sc hvc}\,111$-$07$-$466 are examples of low $|b|$ CHVCs at
velocities unambiguously forbidden in terms of normal galactic
rotation.  The lines of sight in the directions of each of these
features traverse some tens of kpc of the disk before exiting the
Milky Way: unless one is prepared to accept these HVCs as boring
through the conventional disk (for which there is no evidence), and
atypical in view of the cleanliness of the terminal--velocity locus,
then their distance is constrained to be large.

\subsubsection{CHVCs near the galactic poles}

Similarly, HVCs located near the galactic poles offer unambiguous
information on the vertical, $z$, component of their space motion.
Burton (\cite{burt97}) noted that the vertical thickness of
the galactic \hi layer is well measured with a scale height of $h_{\rm
  z} \sim 100$~pc, and so consequently the HVCs either do not commonly
populate the lower galactic disk/halo transition region or else the
$z$ component dominates the total space motion, which seems unlikely
and is contradicted by the examples in the preceding paragraph.
Examples of high vertical velocities from our tabulation include {\sc
  hvc}\,119$-$73$-$301, {\sc hvc}\,148$-$82$-$258, and {\sc
  hvc}\,220$-$88$-$265.  These examples are from the southern galactic
hemisphere, with negative velocities.  The distribution of $z$ motions
over the entries in the tabulation are clearly skewed to negative
velocities.

\subsubsection{CHVCs in the Milky Way cardinal directions}

HVCs in the cardinal directions are interesting because some
ambiguities are removed in these special cases.  For example, {\sc
  hvc}\,087$+$03$-$289, located in the direction of the solar motion
partaking in galactic rotation, has a velocity unambiguously forbidden
for any object within the Milky Way.  The substantial velocities of
{\sc hvc}\,173$-$60$-$236, and {\sc hvc}\,358$+$12$-$137, located
(albeit at substantial latitudes) in directions perpendicular to the
vector of galactic rotation, are also unambiguously forbidden.

\subsection{Completeness and homogeneity of the sample} 

It is appropriate to consider what selection effects might play a role
in the statistics derived from the material in Table~\ref{tab:T_data}.  We
comment below on the relevance of the observational parameters of
sensitivity, velocity coverage, and spatial coverage.

The most stringent of our criteria was the requirement of independent
confirmation.  The Leiden/Dwingeloo survey does, however, involve so
much data that confirming every 5--$\sigma$ spike would have taken an
investment of telescope time which we were not able to make.  There
are features, even in the integrated \hi images shown here, which we have no
reason to consider spurious, but which we simply have not yet
confirmed: in Figure~\ref{fig:F_inthi} examples are seen near $l,b$ =
$86.\!^\circ 5, +1.\!^\circ 0$ in the moment map of {\sc
  hvc}\,087$+$03$-$289; and near $l,b$ = $43.\!^\circ 0, -30.\!^\circ
0$ in the moment map of {\sc hvc}\,039$-$31$-$265; both of these
examples of not--yet--confirmed features are unambiguously in the HVC
regime, and stem from spectra not observed contiguously. (The
observing strategy involved stepping $0.\!^\circ 5$ in $l$, at a
constant $b$, over a $5^\circ$ interval of longitude.) We see no
reason to expect that these features, and many others, could not be
confirmed.

Our use of the terms `compact' and `isolated' is somewhat subjective,
as can be gauged by inspection of the integrated \hi images.
`Compact' is the simpler concept since we have used the somewhat arbitrary
definition of a maximum mean angular size (averaged over the major and
minor axis of elongated features) of 2$^\circ$ {\sc fwhm}. The choice
of a 2$^\circ$ limit was motivated by what appeared to be a natural
break point in the size distribution of HVC features cataloged by
WW91. `Isolated' refers, of course, to the sensitivity level of the
currently available data.  It is not ruled out that some of the
features which we tabulate as isolated in $(l,b,v)$ space would be shown
under scrutiny of more sensitive data to be embedded in a weaker
envelope, or even to be part of a large, but relatively weak, complex
or stream.  For example, entries 47, 49, and 50 in our table, adjacent
on the sky and at comparable velocities, might prove blended under
deeper scrutiny.

The well--known HVC complexes show a range of structures, and within a
complex there are certainly knots of enhanced emission (see Wakker \&
Schwarz \cite{wakke91}).  We did not accept such knots, and so could
plausibly have discriminated against a compact, isolated HVC which
happens to lie in projection against an unrelated extended HVC stream.

Although the Leiden/Dwingeloo survey had been corrected for stray
radiation (Hartmann et al. \cite{hart95}), neither the new Dwingeloo
nor the new Green Bank data were so corrected.  Emission entering the
far--sidelobe pattern is largely contributed by \hi lying, at
relatively modest velocities, in the conventional gaseous disk of the
Milky Way: stray--radiation is not expected over most of the HVC
velocity regime.  Furthermore, stray radiation is contributed from
large solid angles, and thus is diffusely distributed, not
concentrated into point, or very compact, sources like those entering
this discussion.  Therefore we view our results as uncontaminated by
stray radiation.

Selection of candidate CHVCs from the Leiden/Dwingeloo survey was
primarily determined from a significant intensity after smoothing to
32~km~s$^{-1}$ velocity resolution, and a lack of blending.  On the
$0.\!^\circ 5 \times 0.\!^\circ 5$ grid at $\delta \geq -30^\circ$,
unblended clouds (i.e. generally those with a deviation velocity
greater than 50~km~s$^{-1}$) emitting with a 5--$\sigma$ peak
intensity greater than $T_{\rm B} \sim 0.1$~K, and with a {\sc fwhm}
velocity extent broader than 20~km~s$^{-1}$, are unlikely to have been
missed.  Narrow clouds ($<$ 10~km~s$^{-1}$) weaker than about 0.2~K
peak temperature will be underrepresented in the list drawn from the
smoothed Leiden/Dwingeloo data. Although the {\sc fwhm} of most of the
CHVCs tabulated here is greater than 20~km~s$^{-1}$, a few are
considerable narrower, and it is not unexpected that some faint
sources would be missing from our compilation because they were
diluted by a coarse channel spacing. An overall completeness level of
0.2~K in peak brightness is indicated.

The total range of the velocity coverage of the Leiden/Dwingeloo
survey is conservatively quoted as $-450 < v_{\rm LSR} < +400$
km~s$^{-1}$, but in almost all cases it extends usefully ten or more
km~s$^{-1}$ further on both extremes; the total range of the Wakker \&
van Woerden material is larger, extending from $-900$ to $+750$
km~s$^{-1}$.  The HVC with the most extreme velocity known is that
detected by Hulsbosch (\cite{huls78}), and further observed by Cohen
\& Mirabel (\cite{cohe79}) and by Wright (\cite{wrig79}), tabulated
here as {\sc hvc}\,111$-$07$-$466: its $v_{\rm LSR}$ is $-466$
km~s$^{-1}$.  Although a few other HVCs are known with velocities less
than $-400$ km~s$^{-1}$, none is known with comparably extreme
positive velocities.  Therefore it seems reasonable to expect that the
tabulation is not incomplete as a consequence of the velocity range of
the observational material.

In the second stage of preparing our list of CHVC candidates, we used
the Wakker \& van Woerden compilation as input.  The angular lattice
size of that material is larger than that of the Leiden/Dwingeloo
survey.  At $\delta < -18^\circ$, the angular spacing of the Bajaja et
al. (\cite{baja85}) data is $2^\circ \times 2^\circ$. At $\delta >
-18^\circ$, the latitude interval of the Hulsbosch \& Wakker
(\cite{huls88}) data is $1^\circ$ throughout, but the longitude
interval varies from $1^\circ$ at $|b|<45^\circ$ to larger values at
larger $|b|$, while maintaining approximately $\Delta l = 1^\circ$ in
true--angle spacing.  Even though the Leiden/Dwingeloo survey is not
fully sampled to the Nyquist level, the $0.\!^\circ 5$ sampling
interval with a $36'$ beam renders it unlikely that a compact HVC, at
the intensity level being considered here (0.2~K peak brightness
temperature), would have escaped notice in that data because of
undersampling. On the other hand, a compact HVC in our tabulation
could well remain undetected in the Hulsbosch \& Wakker and Bajaja et
al. data as a consequence of the relatively coarse grid spacing.
Hulsbosch \& Wakker (\cite{huls88}) estimate that their material is
essentially complete for point--source clouds with a central
brightness temperature greater than $0.05$ K which lie on an
observational grid point, but only about 57\% complete for
point--source clouds with a central brightness temperature of $0.2$ K
which might not lie on an observational grid point. Insofar as the
WW91 $N \leq 3$ sources served as input for subsequent confirmation,
there may be some additional incompleteness in the strip between
$\delta = -30^\circ$ and $-18^\circ$, where the Bajaja et al.  data
represent a coarser ($2^\circ \times 2^\circ$) sampling and a somewhat
less sensitive detection limit than that of Hulsbosch \& Wakker. 

The above considerations indicate why some CHVCs (23 of the 66) were
found in the Leiden/Dwingeloo data but not in the data used by Wakker
\& van Woerden. It is also possible that a weak CHVC, fortuitously
located in the relevant parameter space of the observations, would be
detected in the Wakker \& van Woerden data but not in the
Leiden/Dwingeloo.  However, any incompleteness due either to
sensitivity, or to lattice size, or to velocity increment, would be
approximately the same everywhere on the observed sky, and so would
not render the tabulation made here unrepresentative of the compact
HVCs.

Other selection effects are plausibly more systematic in specific
velocity ranges and in location.  Because the compilation involved
Dwingeloo data in all cases, either for the initial identification or
for the confirmation, or for both, material is missing where the sky
is not accessible from the Netherlands.  The boundary at $\delta \leq
-30^\circ$ is indicated by the distorted oval drawn in
Figure~\ref{fig:F_lb}.  In view of what is known about the distribution of
high--velocity clouds in the deep southern hemisphere, it seems
reasonable to expect that the southern CHVCs which have been missed
would likely be predominantly at positive radial velocities.  We note
that all 6 of the Local Group galaxies known in the zone $\delta \leq
-30^\circ$ have positive $v_{\rm LSR}$.

Members of the ensemble of compact, isolated HVCs will also be missed
in spectral regions where substantial blending occurs due to
foreground or background emission.  Thus a CHVC with \hi properties
similar to those typical of the average listing in the table would
remain undetected if the emission occurred blended with one of the
major HVC complexes.  Murphy, Lockman, \& Savage (\cite{murp95})
estimate that some 18\% of the sky is covered, at some velocities, by
HVCs to the limit of the Hulsbosch \& Wakker (\cite{huls88}) data.
The major HVC complexes are responsible for such a high areal filling
factor; the areal filling factor for the CHVCs tabulated here is very
small.

The matter of blending becomes more serious at the velocities spanned
by the conventional gaseous disk of the Milky Way, because there the
areal filling factor reaches 100\%.  Although all high--velocity
clouds will have large space velocities, only the line--of--sight
component of the motion is observed.  An HVC could have a space
velocity characteristic of the phenomenon, but an observed $v_{\rm
  LSR}$ of zero.  Some known HVCs trespass on the kinematic regime of
IVCs: for example, Complex C can be traced from the HVC regime into
the kinematic regime associated with IVCs.  The impressive spatial and
kinematic continuities of the Magellanic Stream allow it to be
followed as it traverses from highly deviant positive velocities to
emerge at deviant negative velocities, even though it is lost due to
blending as it crosses the kinematic regime of the conventional Milky
Way gaseous disk, with radial velocities near zero.

But the additional information given by the spatial and kinematic
continuities of a major HVC complex do not pertain for a compact HVC.
If a CHVC were to emit near $v_{\rm LSR} = 0$ km~s$^{-1}$ (and if its
total velocity width were not exceptionally large) then it will have
gone unnoticed and will likely remain so.  Thus deviation velocity is
an incomplete discriminant, although a large deviation velocity is
sufficient cause for considering an emission packet as an HVC, unless
it can be separately demonstrated that emission is contributed from
something else, e.g. from a galaxy.  Blending thus limits the
detection of compact HVCs to those with substantial deviation
velocities.  We comment further below on the possible consequences of
this limitation.

We conclude that the principal causes for incompleteness of our sample
are not systematic in the sense of discriminating against a particular
portion of the sky, except for declinations less than $-30^\circ$ and,
to a much lesser extent, $\delta < -18^\circ$, or in the sense of
being kinematically incomplete, except for CHVCs which might have
trespassed into the low--deviation regime at $|v_{\rm LSR}| < 50$
km/s.  With these caveats, we view the sample as representative of the
class of CHVC objects.

\section{Discussion}
\label{sec:disc}

We compare below the spatial and kinematic deployment of our CHVC
sample with that of the galaxy members of the Local Group as compiled
by Grebel (\cite{greb97}).  The Local Group membership, as currently
estimated, is certainly representative, even if there may be concerns
with the completeness of the Local Group galaxy compilation somewhat
analogous to those pertaining for the CHVC ensemble.  It is not
implausible that additional small galaxies of very low optical surface
brightness will be found in the future; it is also not implausible
that some Local Group galaxy, with brightness comparable to those
already known, remains obscured by the Milky Way, either by dust
extinction or by a high density of foreground stars.  Another analogy
may be drawn between the Local Group galaxy situation and the
situation pertaining to the CHVCs, namely regarding the influence of
the massive Local Group galaxies on the kinematics of objects lying
near M31 or the Milky Way. The LMC, SMC, and the Sagittarius dwarf
spheroidal, are in orbits dominated by the Milky Way, and M32 and
other systems are in orbits dominated by M31, while other Local Group
galaxies may be relatively isolated and distant from either M31 or the
Milky Way, and may have experienced different evolutionary histories.
If the CHVCs pervade the Local Group, a similar discrimination may
pertain.

Fig.~\ref{fig:F_lb} shows the distribution on the sky of the compact,
isolated HVCs, plotted as filled circles.  The locations of galaxies
comprising the Local Group (Grebel \cite{greb97}) are plotted as open
circles.  The sky distribution of CHVCs shows a rather uniform
deployment; in particular, CHVCs do not show the preference for the
northern galactic hemisphere which the total--flux HVC distribution
does, nor do the CHVCs show any tendency to cluster in streams or
complexes.

Our primary concern, of course, is to seek information on the
characteristic distance of the class of compact high--velocity clouds,
as the values of most of the principal physical parameters depend on
distance.  The additional information which is available for a
spatially resolved sample is not available here.  Thus, Blitz et al.
(\cite{blit98}) were able to use the angular size of extended, quite
well resolved HVCs and HVC complexes to estimate distances in a
statistical manner: nearer clouds would, on average, have a larger
angular extent than more distant ones.  Certain other important
kinematic information is also lacking for our sample of generally
unresolved sources, such as that pertaining to kinematic gradients
across an extended feature due, for example, to rotation or to shearing.
Little information is yet available which might reveal relevant
details of the spatial structure of the CHVCs, except for the two
entries, {\sc hvc}\,111$-$07$-$466 and {\sc hvc}\,114$-$10$-$430, for
which Wakker \& Schwarz (\cite{wakke91}) obtained WSRT \hi
observations. It is striking that in both of those cases, the CHVCs
are resolved into elliptical distributions of moderate column density
embedded in diffuse envelopes of low column density. Each elliptical
concentration shows a velocity gradient along the direction of maximum
elongation, rather suggestive of rotation in a flattened disk system.
Further \hi interferometric observations of the CHVC phenomenon which
might, for example, reveal a head--tail structure suggestive of
passage through the halo of the Milky Way or through an intergalactic
medium have not yet been obtained.

We note that because the CHVCs subtend such small angles it will be
particularly difficult to find suitable probes for optical or UV
absorption--line searches toward sources at known distances, thus
possibly, in the most direct way, constraining, or even determining,
the distance to a CHVC. It will similarly be particularly difficult to
establish the metallicity of a CHVC.

As often done in astronomy, and certainly in \hi work, we resort to
the systemic velocities to support discussion of distances
characteristic of the CHVC ensemble.  Fig.~\ref{fig:F_lv} shows the
kinematics of the ensemble of compact, isolated HVCs and of the
ensemble of Local Group galaxies, plotted against galactic longitude
for four different reference frames.

In the upper panel of Fig.~\ref{fig:F_lv}, the longitude distribution
of the motions of the CHVCs is shown, as filled circles, measured with
respect to the Local Standard of Rest, as listed in column 8 of the
table.  Also plotted, as open circles, is the velocity/longitude
distribution of the galaxies in the Local Group, from Grebel's
(\cite{greb97}) catalog.  It is appropriate to comment here on
possible selection effects which might cause systematic distortions to
the true distribution of CHVCs.  The $\delta = -30^\circ$ declination
limit of the LDS \hi observations, as indicated in
Fig.~\ref{fig:F_lb}, is likely to have discriminated against some
CHVCs at positive $v_{\rm LSR}$.  This discrimination will remain in
the other panels in Fig.~\ref{fig:F_lv}, representing velocities in
different reference frames.  We might take the Small Magellanic Cloud
as a test particle illustrative of the missing CHVCs, since it lies
most central, of the Local Group galaxies, in the $\delta = -30^\circ$
oval plotted in Fig.~\ref{fig:F_lb}. The SMC may be followed in the
Fig.~\ref{fig:F_lv} panels, as the open circle near $l=303^\circ$.
Its velocity drops from $+$149 km/s in the $v_{\rm LSR}$ frame, to
near zero velocity in the $v_{\rm GSR}$ and $v_{\rm LGSR}$ frames, and
it is plausible to predict that not--yet--detected CHVCs would follow
the same general tendency.  There may further be a mild discrimination
against detecting CHVCs due to their submersion in the ``\hi zone of
avoidance", i.e. near $v_{\rm LSR} = 0$ \kms, with the velocity
distribution slightly skewed to negative $v_{\rm LSR}$ in galactic
quadrant II, and slightly skewed to positive velocities in quadrant
III.

In the second panel from the top of Fig.~\ref{fig:F_lv}, the motions of
the CHVC ensemble and of the Local Group galaxies are plotted with
respect to the Galactic Standard of Rest; and in the third panel, with
respect to the Local Group Standard of Rest.  The adopted definitions
of the various velocity systems with units of km~s$^{-1}$ are: 

\begin{equation}
v_{\rm LSR}=v_{\rm HEL}+9\cos(l)\cos(b)+12\sin(l)\cos(b)-7\sin(b)
\end{equation}
\begin{equation}
v_{\rm GSR}=v_{\rm LSR}+0\cos(l)\cos(b)+220\sin(l)\cos(b)+0\sin(b)
\end{equation}
\begin{equation}
v_{\rm LGSR}=v_{\rm GSR}-62\cos(l)\cos(b)+40\sin(l)\cos(b)-35\sin(b)
\end{equation}

Fig.~\ref{fig:F_vhist} shows histograms of the velocities in these
reference frames; the dispersion of the velocities decreases in a
progression from the $v_{\rm LSR}$ reference frame, for which
$\sigma_{\rm LSR} = 175$ km~s$^{-1}$, via the $v_{\rm GSR}$ one
($\sigma_{\rm GSR} = 95$ km~s$^{-1}$), to the $v_{\rm LGSR}$ frame,
for which $\sigma_{\rm LGSR} = 88$ km~s$^{-1}$.  As noted
by Blitz et al. (\cite{blit98}) a decreasing velocity dispersion for a
population gives a good indication that a more appropriate reference
frame is being approached. They cite the example of the globular
cluster system of the Galaxy, for which the velocity dispersion drops
from 134 to 119~km~s$^{-1}$ in going from the LSR to the more relevant
GSR frame.  

Since our CHVC sample has both a substantial size and an essentially
uniform distribution on the sky (Fig.~\ref{fig:F_lb}) it is appropriate to
use the sample itself to define a best-fitting velocity reference
system. We have determined the direction cosine
coefficients that provide a minimum velocity dispersion of the
measured radial velocities. The result of this optimization has been
labeled the ``XSR system'' and is defined by:

\begin{equation}
v_{\rm XSR} = v_{\rm GSR}+0\cos(l)\cos(b)+45\sin(l)\cos(b)-90\sin(b)
\end{equation}

The velocity dispersion of the CHVC sample in the XSR system is only
$\sigma_{\rm XSR}=69$ km~s$^{-1}$. The accuracy of the coefficients of
the direction cosines, and therefore the implied solar apex
$(l_\odot,b_\odot,v_\odot) = (88^\circ,-19^\circ,+293$~km~s$^{-1}$),
is, however, not very high. Varying each coefficient by plus and minus
50~km~s$^{-1}$ increases the dispersion of the distribution from its
minimum value to about 75~km~s$^{-1}$. Comparable uncertainties of
perhaps 50~km~s$^{-1}$ also apply to the coefficients which define the
LGSR frame, $(l_\odot,b_\odot,v_\odot) =
(93^\circ,-4^\circ,+316$~km~s$^{-1}$) according to Karachentsev \& Makarov
(\cite{kara96}), implying agreement between these two frames at about
the one sigma level. This is particularly interesting since the XSR
system was defined completely independently on the basis of the CHVC
system alone.

Figure~\ref{fig:F_cost} shows the heliocentric velocity plotted against
the cosine of the angle between the solar apex and the direction to
the CHVCs, indicated by closed circles, and the directions to the
Local Group galaxies, indicated by open circles.  Such displays of the
kinematics have been used to ascertain membership of galaxies in the
Local Group by Sandage (\cite{sand86}), van den Bergh
({\cite{berg94}), Grebel (\cite{greb97}), and others. The dashed lines
at $\pm 60$~km~s$^{-1}$ represent the 1--$\sigma$ envelope
considered by Sandage (\cite{sand86}) to describe the dispersion of
the Local Group membership.  Van den Bergh (\cite{berg94}) notes
that galaxies on the outer fringe of the Local Group, about 1~Mpc
distant from the Local Group barycenter, tend to lie above the upper
envelope.  Galaxies with quite large positive--velocity deviations
would be partaking in the cosmological expansion, and at distances
substantially greater than 1~Mpc.  We note that the filled circle
most deviant from the mean of the CHVC objects is the galaxy
interloper, Cepheus~1, which Burton et al.  (\cite{burt99}) judge,
mainly by its angular and kinematic proximity to NGC~6946, to be at
a distance of approximately 6~Mpc.  The most deviant of the Local
Group galaxies is the Sagittarius dwarf spheroidal, which lies so
close to the Milky Way that its motion would be gravitationally
distorted and thus not representative of the Local Group ensemble.
The position of Cepheus~1 ($l=94^\circ$) and of the Sagittarius
dwarf spheroidal ($l=6^\circ$) are also, not surprisingly,
unrepresentative of the means in the velocity/longitude plots of
Fig.~\ref{fig:F_lv}.

Despite the similarities between the deployment of the CHVCs and the
Local Group galaxies in Figs.~\ref{fig:F_lv} and \ref{fig:F_cost}, there is a
clear tendency for the distribution of CHVC velocities to have a
negative mean.  This same tendency is evident at most galactic
longitudes and in all reference frames, but is of course clearest in
the LGSR and XSR frames for which a net infall of the CHVCs with a
mean velocity of about 100~km/~s$^{-1}$ is measured.  This point needs
some further investigation in terms of the selection criteria of the
CHVC sample, in particular to gauge the extent to which undetected
CHVCs with $v_{\rm LSR}$ near zero might influence the result.
Inspection of the LSR velocity histogram in Fig.~\ref{fig:F_vhist}
suggests that perhaps a total of 8 objects might be missing in this
velocity range from an otherwise continuous distribution. It seems
unlikely that inclusion of the ``missing'' objects would dramatically
influence the result.  Taken at face value, these kinematics are
suggestive of a population which is bound to the Local Group, but
which has not yet experienced significant interaction or merger with
the larger members, which would tend to virialize their motions. A
mean radial infall velocity which is comparable to the velocity
dispersion of the Local Group member galaxies, $\sigma_{\rm LGSR} =
76$ km~s$^{-1}$, is quite plausible if the same gravitational
potential is responsible for both.  Perhaps of relevance is the fact
that C\^ot\'e et al.  (\cite{cote97}) found that the faint dwarf
galaxies in the two groups of galaxies nearest to the Local Group show
a wider range of velocity distribution than the brighter members. An
evolutionary history which is a function of mass may be a natural part
of the galaxy formation process.

\setlength{\tabcolsep}{2pt}
\begin{flushleft} 
\begin{table*} 
\caption[]{Compact, isolated high-velocity clouds.} \label{tab:T_data}
\rotate{
\begin{tabular}{rclrrrrrrrrccccl} 
\noalign{\smallskip} \hline 
\# & \hspace{.4cm} designation & data & $l$~~ & $b$~~ & RA~ & Dec & $v_{\rm LSR}$ 
& $v_{\rm GSR}$ 
& $T_{\rm max}$ & {\sc FWHM} & $\int {\rm T_B}dV$ & Maj & Min & PA & references \\ 
\ & HVC~$lll \pm bb \pm vvv$ & I/C &($^\circ$) &($^\circ$) & (h m) &
($^\circ~\prime$) & (km/s) & (km/s) & (K) & 
(km/s) & (K~km/s) & ($^\circ$) &($^\circ$) &($^\circ$) & \\ 
\noalign{\smallskip} \hline 
 1 & HVC~017$-$25$-$218 & LD/Dm &  16.58 &$-$25.28&  19 59.1& $-$24 55&$-$227 &$-$170 & 0.45 & 13.5 &  6.5 & 0.6 & 0.2 & 20 & \small{M81} \\ 
 2 & HVC~018$+$47$-$145 & LD/Dm &  18.15 &$+$47.01&  15 39.7& $ $10 18&$-$145 &$-$98  & 0.15 & 42.8 &  6.8 & 1.1 & 0.5 & 60 & \small{W57} \\ 
 3 & HVC~024$-$02$-$285 & LD/D  &  24.29 &$-$1.96 &  18 42.6& $-$08 35&$-$284 &$-$194 & 0.32 & 25.8 &  8.7 & 1.1 & 0.5 & 90 & \\ 
 4 & HVC~030$-$51$-$119 & LD/Dm &  29.56 &$-$50.69&  21 58.3& $-$22 42&$-$123 &$-$54  & 0.67 & 39.4 & 28.0 & 1.1 & 0.4 & 95 & \\ 
 5 & HVC~031$-$20$-$287 & LD/D  &  31.38 &$-$20.09&  20 01.0& $-$10 20&$-$287 &$-$179 & 0.42 & 36.0 & 15.9 & 0.8 & 0.5 & 40 & \small{W386} \\ 
 6 & HVC~032$-$31$-$299 & LD/Dm &  31.90 &$-$30.77&  20 41.7& $-$14 18&$-$309 &$-$209 & 0.24 & 31.2 &  7.9 & 0.5 & 0.3 & 90 & \small{W443} \\ 
 7 & HVC~039$-$37$-$231 & LD/Dm &  38.72 &$-$37.31&  21 15.6& $-$11 50&$-$238 &$-$129 & 0.33 & 17.7 &  6.1 & 0.6 & 0.2 & 60 & \small{W482} \\ 
 8 & HVC~039$-$33$-$260 & LD/D  &  38.78 &$-$33.47&  21 01.6& $-$10 08&$-$260 &$-$145 & 0.43 & 26.3 & 12.1 & 0.7 & 0.7 &  0 & \small{W460} \\ 
 9 & HVC~039$-$31$-$265 & LD/Dm &  39.40 &$-$30.67&  20 52.4& $-$08 25&$-$278 &$-$158 & 0.33 & 35.9 & 12.7 & 1.8 & 0.7 &120 & \small{W442} \\ 
10 & HVC~040$+$01$-$282 & W/LD  &  40.20 &$+$0.60 &  19 02.6& $ $06 44&$-$282 &$-$140 & 0.12 & 42.6 &  5.5 & 1.0 & 0.5 & 90 & \small{W289}  \\ 
11 & HVC~043$-$13$-$302 & LD/Dm &  42.91 &$-$13.11&  19 56.3& $ $02 41&$-$314 &$-$168 & 0.53 & 27.6 & 15.6 & 1.0 & 0.7 & 30 & \small{G81,W348} \\ 
   &                    &       &        &        &         & $ $     &$-$267 &$-$121 & 0.22 & 39.7 &  9.3 &     &     &    & \\ 
12 & HVC~047$-$52$-$129 & LD/D  &  47.14 &$-$51.73&  22 19.4& $-$12 52&$-$129 &$-$29  & 0.24 & 25.4 &  6.4 & 2.3 & 2.0 & 90 & \small{W521} \\
13 & HVC~050$-$68$-$187 & LD/Dm &  49.80 &$-$68.28&  23 23.4& $-$19 09&$-$201 &$-$139 & 0.19 & 31.3 &  6.3 & 0.6 & 0.4 & 35 & \\ 
14 & HVC~060$-$65$-$206 & W/LD  &  59.80 &$-$64.60&  23 18.6& $-$13 51&$-$206 &$-$124 & 0.12 & 38.3 &  5.1 & 0.7 & 0.5 & 30 & \small{W545}  \\ 
15 & HVC~069$+$04$-$223 & LD/Dm &  69.16 &$+$3.75 &  19 50.1& $ $33 41&$-$234 &$ $29  & 0.24 & 34.0 &  8.8 & 1.0 & 0.9 & 30 & \\ 
16 & HVC~070$+$51$-$146 & W/LD  &  70.10 &$+$50.80&  15 48.8& $ $43 53&$-$146 &$-$15  & 0.20 & 33.0 &  7.0 & 1.2 & 0.6 & 60 & \small{W44}  \\ 
17 & HVC~087$+$03$-$289 & LD/Dm &  86.82 &$+$2.92 &  20 46.3& $ $47 50&$-$298 &$-$79  & 0.16 & 35.7 &  6.1 & 1.5 & 0.4 &110 & \small{W278} \\ 
18 & HVC~092$-$39$-$367 & LD/D  &  92.47 &$-$39.41&  23 14.4& $ $17 39&$-$367 &$-$197 & 0.21 & 27.0 &  6.0 & 1.6 & 1.2 &115 & \\ 
19 & HVC~094$+$08$+$080 & LD/Dm &  94.40 &$+$7.93 &  20 51.7& $ $56 52&$+$65  &$+$282 &$\sim 0.3$& -- & -- & 0.0 & 0.0 &  0 & \small{BBWH99} \\ 
20 & HVC~095$-$63$-$314 & W/LD  &  94.70 &$-$63.20&  00 02.1& $-$03 04&$-$314 &$-$215 & 0.35 & 26.9 & 10.1 & 1.0 & 0.8 &160 & \small{W542}  \\ 
21 & HVC~100$-$49$-$383 & LD/Dm &  99.58 &$-$48.89&  23 49.8& $ $11 10&$-$388 &$-$245 & 0.21 & 51.8 & 11.8 & 0.8 & 0.5 & 45 & \\ 
22 & HVC~107$-$30$-$421 & LD/GB & 107.49 &$-$29.72&  23 48.7& $ $31 16&$-$419 &$-$237 & 0.18 & 31.5 &  6.2 & 1.2 & 0.5 & 50 & \small{W437,H92} \\ 
23 & HVC~108$-$21$-$388 & LD/GB & 108.10 &$-$21.42&  23 39.6& $ $39 22&$-$395 &$-$200 & 0.11 & 26.1 &  3.1 & 0.6 & 0.4 &110 & \small{W389} \\ 
24 & HVC~111$-$07$-$466 & LD/GB & 110.58 &$-$7.00 &  23 27.1& $ $53 50&$-$466 &$-$262 & 0.17 & 24.9 &  4.6 & 0.8 & 0.7 & 65 & \small{H78,CM79,Wr79,W318} \\ 
25 & HVC~114$-$10$-$430 & LD/GB & 113.56 &$-$10.49&  23 52.1& $ $51 18&$-$440 &$-$242 & 0.77 & 10.0 &  8.2 & 0.4 & 0.3 & 90 & \small{H78,CM79,W330} \\ 
26 & HVC~115$+$13$-$275 & LD/Dm & 115.39 &$+$13.38&  22 56.9& $ $74 33&$-$260 &$-$67  & 0.16 & 95.4 & 16.5 & 1.1 & 0.6 &100 & \\ 
27 & HVC~118$-$58$-$373 & W/LD  & 118.50 &$-$58.20&  00 42.1& $ $04 35&$-$373 &$-$271 & 0.66 & 30.8 & 21.6 & 1.0 & 0.7 &140 & \small{MC79,G81,W532}  \\ 
28 & HVC~119$-$73$-$301 & W/LD  & 118.50 &$-$73.10&  00 46.2& $-$10 16&$-$301 &$-$245 & 0.13 & 31.6 &  4.4 & 0.7 & 0.4 & 80 &  \small{W555}  \\
29 & HVC~119$-$31$-$381 & LD/D  & 119.17 &$-$30.92&  00 36.2& $ $31 50&$-$381 &$-$216 & 0.29 & 24.6 &  7.5 & 0.9 & 0.8 &100 & \small{Wr79,W444,H92} \\ 
30 & HVC~123$-$32$-$324 & W/LD  & 122.90 &$-$32.20&  00 51.3& $ $30 40&$-$324 &$-$168 & 0.16 & 32.7 &  5.4 & 0.7 & 0.4 & 30 & \small{W446}  \\ 
31 & HVC~125$+$41$-$207 & W/LD  & 125.20 &$+$41.40&  12 24.0& $ $75 36&$-$207 &$-$72  & 2.10 &  5.9 & 13.2 & 1.3 & 0.7 &120 & \small{W84}  \\ 
32 & HVC~129$+$15$-$295 & LD/Dm & 128.82 &$+$14.97&  02 33.2& $ $76 40&$-$306 &$-$140 & 0.44 & 18.1 &  8.5 & 0.0 & 0.0 &  0 & \small{W231} \\ 
33 & HVC~133$-$76$-$285 & LD/Dm & 132.80 &$-$75.87&  01 01.3& $-$13 11&$-$296 &$-$257 & 0.16 & 37.9 &  6.3 & 0.6 & 0.1 & 90 & \small{W557} \\ 
\noalign{\smallskip} \hline 
\end{tabular}} 
\end{table*} 
\end{flushleft}
\begin{flushleft} 
\begin{table*} \caption[]{Compact, isolated high-velocity clouds. continued} 
\rotate{
\begin{tabular}{rclrrrrrrrrccccl} 
\noalign{\smallskip} \hline 
\# & \hspace{.4cm} designation & data & $l$~~ & $b$~~ & RA~ & Dec & $v_{\rm LSR}$ 
& $v_{\rm GSR}$ 
& $T_{\rm max}$ & {\sc FWHM} & $\int {\rm T_B}dV$ & Maj & Min & PA & references \\ 
\ & HVC~$lll \pm bb \pm vvv$ & I/C &($^\circ$) &($^\circ$) & (h m) &
($^\circ~\prime$) & (km/s) & (km/s) & (K) & 
(km/s) & (K~km/s) & ($^\circ$) &($^\circ$) &($^\circ$) & \\ 
\noalign{\smallskip} \hline 
34 & HVC~146$-$78$-$275 & LD/D  & 146.01 &$-$77.65&  01 11.4& $-$15 41&$-$275 &$-$249 & 0.16 & 23.9 & 4 .1 & 0.9 & 0.9 &110 & \small{W560} \\ 
35 & HVC~148$-$32$-$144 & LD/Dm & 147.89 &$-$31.92&  02 25.7& $ $26 21&$-$146 &$-$47  & 0.18 & 35.2 &  6.8 & 0.5 & 0.3 & 80 & \\ 
36 & HVC~148$-$82$-$258 & LD/Dm & 148.04 &$-$82.46&  01 05.0& $-$20 16&$-$267 &$-$252 & 0.47 & 20.2 & 10.1 & 0.0 & 0.0 &  0 & \\ 
37 & HVC~149$-$38$-$140 & LD/D  & 149.24 &$-$37.92&  02 19.1& $ $20 26&$-$140 &$-$51  & 0.27 & 29.3 &  8.3 & 1.0 & 0.7 &100 & \\ 
38 & HVC~157$+$03$-$185 & LD/D  & 156.63 &$+$2.73 &  04 46.4& $ $49 35&$-$185 &$-$98  & 0.18 & 20.3 &  3.8 & 1.5 & 1.1 & 90 & \small{W275} \\ 
39 & HVC~158$-$39$-$285 & W/LD  & 157.80 &$-$39.20&  02 41.3& $ $16 07&$-$285 &$-$221 & 0.16 & 37.9 &  6.4 & 1.3 & 0.8 &150 & \small{W486}  \\ 
40 & HVC~162$+$02$-$170 & LD/Dm & 161.87 &$+$2.46 &  05 04.5& $ $45 20&$-$181 &$-$113 & 0.59 & 27.9 & 17.5 & 0.8 & 0.7 & 55 & \small{W277} \\ 
41 & HVC~171$-$54$-$229 & LD/Dm & 170.97 &$-$53.77& 02 35.7& $ $00 55&$-$235 &$-$215 & 0.49 & 23.7 & 12.4 & 1.0 & 0.4 & 80 & \small{H78,W525} \\ 
42 & HVC~172$+$51$-$114 & W/LD  & 171.60 &$+$51.40& 10 00.6& $ $46 18&$-$114 &$-$94  & 0.27 & 38.6 & 11.0 & 3.0 & 1.2 &150 & \small{W39}  \\ 
43 & HVC~173$-$60$-$236 & W/LD  & 172.70 &$-$59.50& 02 23.1& $-$05 48&$-$236 &$-$222 & 0.22 & 17.6 &  4.1 & 0.7 & 0.5 & 90 & \small{W536}  \\ 
44 & HVC~186$+$19$-$114 & W/LD  & 186.20 &$+$18.90& 07 16.9& $ $31 46&$-$114 &$-$136 & 1.03 & 20.4 & 22.4 & 1.0 & 0.8 &140 & \small{W215}  \\ 
45 & HVC~191$+$60$+$093 & LD/GB & 190.86 &$+$60.36& 10 36.9& $ $34 10&$+$93  &$ $73  & 0.38 & 29.6 & 12.0 & 1.0 & 0.9 & 75 & \\ 
46 & HVC~198$-$12$-$103 & W/LD  & 197.70 &$-$11.60& 05 41.8& $ $07 54&$-$103 &$-$169 & 0.48 & 23.9 & 12.2 & 1.0 & 0.8 &110 & \small{W343}  \\ 
47 & HVC~200$+$30$+$080 & LD/Dm & 200.22 &$+$29.72& 08 22.2& $ $23 20&$+$75  &$+$9   & 0.50 & 28.9 & 15.3 & 0.0 & 0.0 &  0 & \\ 
48 & HVC~200$-$16$-$091 & W/LD  & 200.50 &$-$15.70& 05 32.9& $ $03 30&$-$91  &$-$165 & 0.56 & 27.1 & 16.0 & 1.1 & 0.9 &150 & \small{W362}  \\ 
49 & HVC~202$+$30$+$057 & LD/D  & 202.23 &$+$30.38& 08 27.4& $ $21 55&$+$57  &$-$15  & 1.12 & 26.0 & 30.9 & 1.8 & 1.3 & 35 & \\ 
50 & HVC~204$+$30$+$075 & LD/Dm & 204.15 &$+$29.80& 08 27.5& $ $20 09&$+$61  &$-$17  & 1.19 & 33.9 & 42.8 & 0.8 & 0.6 &145 & \\ 
51 & HVC~220$-$88$-$265 & LD/D  & 219.75 &$-$88.06& 01 00.1& $-$27 21&$-$265 &$-$270 & 0.30 & 12.8 &  4.2 & 1.5 & 1.3 & 90 & \\ 
52 & HVC~224$-$08$+$192 & W/LD  & 224.20 &$-$8.00 & 06 43.0& $-$13 38&$+$192 &$+$40  & 0.24 & 26.3 &  6.8 & 1.2 & 0.9 & 90 & \small{W325}  \\ 
53 & HVC~225$+$36$+$082 & W/LD  & 224.60 &$+$35.90& 09 19.2& $ $06 59&$+$82  &$-$43  & 0.22 & 36.3 &  8.6 & 1.3 & 1.0 & 80 & \small{W115}  \\ 
54 & HVC~225$-$42$+$190 & LD/Dm & 224.87 &$-$41.60& 04 29.8& $-$26 08&$+$184 &$+$68  & 0.22 & 35.8 &  8.2 & 0.8 & 0.5 & 95 & \\ 
55 & HVC~227$-$34$+$114 & LD/Dm & 226.84 &$-$33.57& 05 06.0& $-$25 30&$+$115 &$-$19  & 0.60 & 32.3 & 20.7 & 0.8 & 0.4 & 90 & \\ 
56 & HVC~229$-$74$-$168 & LD/Dm & 228.93 &$-$74.22& 02 02.0& $-$30 22&$-$174 &$-$219 & 0.25 & 33.4 &  8.8 & 0.7 & 0.5 &120 & \\ 
57 & HVC~230$+$61$+$165 & LD/GB & 230.36 &$+$60.62& 10 55.2& $ $15 28&$+$155 &$+$72  & 0.23 & 29.3 &  7.1 & 1.4 & 0.7 &120 & \\ 
58 & HVC~235$-$74$-$150 & LD/Dm & 235.26 &$-$73.74& 02 02.7& $-$32 10&$-$157 &$-$208 & 0.27 & 23.1 &  6.6 & 0.7 & 0.5 & 80 & \\ 
59 & HVC~237$+$50$+$078 & W/LD  & 236.70 &$+$49.80& 10 25.4& $ $06 42&$+$78  &$-$41  & 0.19 & 36.8 &  7.6 & 1.2 & 1.2 &  0 & \small{W47}  \\ 
60 & HVC~241$+$53$+$089 & W/LD  & 241.00 &$+$53.40& 10 43.5& $ $06 41&$+$89  &$-$26  & 0.15 & 59.7 &  9.8 & 1.2 & 0.8 & 50 & \small{W34}  \\ 
61 & HVC~263$+$27$+$153 & W/LD  & 263.00 &$+$26.90& 10 16.6& $-$23 43&$+$153 &$-$42  & 1.10 & 22.2 & 26.4 & 1.3 & 0.8 & 40 & \small{W162}  \\ 
62 & HVC~267$+$26$+$216 & W/LD  & 267.30 &$+$26.10& 10 28.1& $-$26 41&$+$216 &$+$19  & 2.61 & 19.0 & 52.8 & 1.6 & 1.0 & 10 & \small{GH77,W176}  \\ 
63 & HVC~271$+$29$+$181 & W/LD  & 271.50 &$+$28.90& 10 48.9& $-$26 23&$+$181 &$-$12  & 0.92 & 24.1 & 23.6 & 1.3 & 0.8 &110 & \small{W163}  \\ 
64 & HVC~284$-$84$-$174 & W/LD  & 284.00 &$-$84.00& 01 00.7& $-$32 47&$-$174 &$-$196 & 0.35 & 28.6 & 10.5 & 1.2 & 1.0 &  0 & \small{W561}  \\ 
65 & HVC~340$+$23$-$108 & LD/D  & 340.12 &$+$22.51& 15 29.6& $-$28 43&$-$108 &$-$177 & 0.36 & 32.4 & 12.5 & 1.2 & 0.5 & 90 & \\ 
66 & HVC~358$+$12$-$137 & LD/D  & 358.11 &$+$12.29& 16 55.6& $-$23 33&$-$137 &$-$144 & 0.56 & 19.4 & 11.5 & 1.1 & 0.7 &120 & \\ 
\noalign{\smallskip} \hline 
\end{tabular} }
\end{table*}
\end{flushleft} 

\section{Summary} 

We have attempted to determine an objectively defined class of
high--velocity clouds, which might represent a homogeneous subsample
of these objects, in a single physical state.  The selection criteria
led to a catalog of compact, isolated high--velocity clouds; the
criteria excluded the Magellanic Stream and all of the other known HVC
complexes from our considerations.  We are, of course, aware that the
total \hi flux represented by all of the CHVCs is less than that from
a single large complex.  A full explanation of the HVC phenomenon will
have to subsume the complexes as well as the compact objects, and it
is not yet clear if a single, unifying explanation will suffice.

No direct distance determination is yet available for any of the
objects.  Several aspects of the topology of the class are difficult
to account for if the CHVCs are viewed as a Milky Way population, in
particular if they are viewed as consequences of a galactic fountain.
The amplitude of the horizontal motions of these ``bullets" is
comparable to that of the vertical motions.  The vertical motions are
larger than expected for free fall onto the Milky Way from material
returning in a fountain flow.  There is no preference shown for the
terminal--velocity locus, where motions from violent events leading to
a fountain would be expected to be most common.  Unlike the situation
if the major HVC complexes are considered, the CHVCs are scattered
rather uniformly across the sky, with no strong preference for the
northern galactic hemisphere.  The CHVCs show no tendency to
accumulate in the lower halo of the Milky Way.  The CHVCs also show no
tendency to cluster along filamentary structures.  Regarding both
their spatial and kinematic distributions, the CHVCs show substantial
similarities with the distributions of the galaxies comprising the
Local Group. The solar apex which follows directly from a minimization
of the velocity dispersion of the CHVC system namely,
$(l_\odot,b_\odot,v_\odot) = (88^\circ,-19^\circ,+293$~km~s$^{-1}$),
agrees within the errors with that which defines the Local Group
Standard of Rest, $(l_\odot,b_\odot,v_\odot) =
(93^\circ,-4^\circ,+316$~km~s$^{-1}$), found by Karachentsev \& Makarov
(\cite{kara96}). The velocity dispersion of the CHVC system in this
reference frame is only $\sigma_{\rm XSR}=69$~km~s$^{-1}$, while there
is a mean infall of $v_{\rm LGSR}=v_{\rm XSR}=-100$~km~s$^{-1}$. 

It seems that the most plausible reference system for the CHVC
deployment and kinematics is that of the Local Group. The low velocity
dispersion in this reference frame and substantial radial infall are
strongly suggestive of a population which has as yet had little
interaction with the more massive Local Group members, which have both
a slightly larger velocity dispersion $\sigma_{\rm
  LGSR}=76$~km~s$^{-1}$ and a small positive mean velocity $v_{\rm
  LGSR}=+22$~km~s$^{-1}$. At a typical distance of about 1~Mpc the
CHVCs would have sizes of about 15~kpc and gas masses, M$_{\rm HI}$, of
a few times 10$^7$M$_\odot$, corresponding to those of (sub-)dwarf
galaxies. Although the total gas mass represented by our entire sample would
not be large, M$_{\rm HI}$, of a few times 10$^9$M$_\odot$, it is still
quite substantial. More importantly, the CHVCs may still represent 
pristine examples of collapsed objects, with only a small amount of
internal star formation and enrichment. As such, they should provide
substantial insight into the process of galaxy and structure formation.

Since the inception of high--velocity cloud research, the possibility
of an extragalactic deployment of these clouds has been critically
considered as a possibility; among others, the discussions by Oort
(\cite{oort66}, \cite{oort70}, \cite{oort81}), Verschuur
(\cite{vers75}), Giovanelli (\cite{giov81}), Bajaja, Morras, \&
P\"oppel (\cite{baja87}), Wakker \& van Woerden (\cite{wakk97}), and
Blitz et al. (\cite{blit98}) are particularly relevant here.  The role
played by the HVC complexes is important to each of these discussions.
Giovanelli (\cite{giov81}) tried, as we have done here, to define an
appropriately unambiguous subsample of HVCs, which avoided the
complexes: his subsample of HVCs with extreme (negative) velocities
lay in one galactic quadrant, at $l<180^\circ$ and $b<0^\circ$.
Giovanelli considered it more likely that the HVCs in his sample were
shreds of the Magellanic Stream precipitating toward the galactic
disk.  We note that the sample of CHVCs considered here covers much of
the sky, and includes members far removed in both angle and velocity
from the Magellanic Stream; we note also that the Local Group galaxies
with the most extreme (negative) velocities -- see Fig.~\ref{fig:F_lv} -- also
populate that single quadrant.  Blitz et al. (\cite{blit98}), who also
remove the Magellanic Stream and several major complexes from their
discussion, justify this distinction as not arbitrary, but supported
by observational constraints on the distances of the excluded
complexes.  They conclude that the remaining HVC material, including a
substantial flux residing in the remaining complexes, is falling onto
the Local Group.  Blitz et al. connect the HVC properties to the
hierarchical structure formation scenario, and to the gas seen as
Lyman--limit clouds in absorption towards quasars; some of their
predictions and considerations are also relevant for the CHVCs.

Several aspects of the compact, isolated high--velocity objects seem
particularly appropriate for further observations.  Many of the CHVCs
cataloged are suitable candidates for \hi radio interferometry.  The
CHVCs show such a wide range of observed linewidths that a number of
questions arise concerning stability and possible rotation.  It would
be interesting to see if any of the CHVCs showed internal kinematics,
particularly rotation, at an amplitude which would indicate the
presence of dark matter, or which might suggest that the objects
resemble dwarf galaxies in which stars have not yet been found, or
have simply not formed.  Although it seems that optical or UV
absorption--line studies will be hindered by a paucity of suitable
background probes, many of the CHVCs are suitable candidates for deep
probes of optical emission.  If the CHVCs are at Local Group
distances, the diffuse H$\alpha$ emission surrounding the CHVCs would
be less bright than that associated with the HVCs in complexes lying
in the halo of the Milky Way. However, the H$\alpha$ emission from
even a single \hii region or Planetary Nebula could be easily
detected, and this would allow immediate recognition of an associated
stellar population and hence a distance.  Deep optical probes would
help clarify the distinction between the CHVCs and nearby dwarf
galaxies, and would, at least, reveal any remaining major interloper
like Cepheus~1 which we found masquerading in our catalog as {\sc
  hvc}\,094$+$08$+$080.

\begin{acknowledgements} 
  We are grateful to B.\,P. Wakker for providing us with a digital
  version of the (updated) HVC catalogue of Wakker \& van Woerden
  (\cite{wakk91}), and to E.\,K. Grebel for providing a digital
  version of her catalogue of members of the Local Group. The
  Dwingeloo radio telescope is operated by the Netherlands Foundation
  for Research in Astronomy, under contract with the Netherlands
  Organization for Scientific Research.  The Green Bank 140--foot
  telescope is operated by the National Radio Astronomy Observatory
  under contract with the U.\,S. National Science Foundation; we are
  grateful to F.\,J. Lockman and D. Balser for obtaining some of the
  Green Bank spectra at our request.
\end{acknowledgements} 

{} 

\begin{figure} 

\caption{Images of integrated \hi emission, paired with a 
  representative spectrum, for each of the 66 compact, isolated HVCs
  tabulated in our sample.  \hi emission is integrated over a velocity
  extent of 200 km~s$^{-1}$, centered approximately on the mean
  velocity of the CHVC.  The data were extracted from the
  Leiden/Dwingeloo survey CD--ROM.  The associated spectrum refers to
  the direction in the $0.\!^\circ 5 \times 0.\!^\circ 5$ grid nearest
  to the peak of the integrated emission.  For those compact HVCs
  confirmed with new Green Bank data (as indicated in column 3 of the
  table), the spectrum displayed was obtained on the 140--foot
  telescope; for all other entries, the spectrum displayed is from the
  Leiden/Dwingeloo survey.  } \label{fig:F_inthi}
\end{figure} 

\begin{figure} 

\caption{Distribution on the sky of the compact, isolated HVCs, plotted as 
  filled circles.  The sample of CHVCs is unlikely to be complete, but
  it is, as discussed in the text, homogeneous and probably
  representative of this class of object.  Unlike the major HVC
  complexes, the CHVCs are distributed rather uniformly over the sky.
  The sky distribution of Local Group galaxies (from Grebel
  \cite{greb97}) is shown by open circles.} \label{fig:F_lb}
\end{figure} 

\begin{figure} 
 
\caption{Velocities of the ensemble of compact, isolated HVCs and of the 
  ensemble of Local Group galaxies, plotted against galactic longitude
  for four different kinematic reference frames. The CHVCs are shown
  as filled circles; the members of the Local Group, as open circles.
  In the upper panel, the motions of the CHVCs are shown measured with
  respect to the Local Standard of Rest; in the second panel from the
  top, with respect to the Galactic Standard of Rest; and in the third
  panel, with respect to the Local Group Standard of Rest.  Figure 4
  shows histograms of the velocities in these reference frames; the
  dispersion of the velocities decreases in a progression from the
  $v_{\rm LSR}$ reference frame, via the $v_{\rm GSR}$ one, to the
  $v_{\rm LGSR}$ frame.  The bottom panel here shows the CHVC motions
  measured with respect to a reference frame, labeled XSR, which
  minimizes the dispersion of the motions.  } \label{fig:F_lv}
\end{figure} 

\begin{figure} 

\caption{Histograms of distributions of the CHVC velocities, and of Local 
  Group galaxy velocities, as measured in the reference frames
  indicated on the absciss\ae.  The open histograms represent the CHVC
  ensemble; the hatched histograms, the Local Group galaxies.  The dispersion
  of the histogram representing the $v_{\rm LGSR}$ frame is
  significantly smaller than that in the $v_{\rm LSR}$ and $v_{\rm
    GSR}$ reference frames.  The velocities labelled $v_{\rm XSR}$
  refer to a frame which was found to minimize the dispersion of the
  CHVC velocity distribution. } \label{fig:F_vhist}
\end{figure} 

\begin{figure} 

\caption{Variation of heliocentric velocity versus the cosine of the 
  angular distance between the solar apex and the $(l,b)$ direction of
  the object.  Members of the compact, isolated high--velocity cloud
  ensemble are plotted as filled circles; galaxies comprising the
  Local Group, as open circles.  The solid line represents the solar
  motion of $v_\odot = 316$ km~s$^{-1}$ toward $l=93^\circ$,
  $b=-4^\circ$ as determined by Karachentsev \& Makarov
  (\cite{kara96}) and used by Grebel (\cite{greb97}) in her analysis.
  The dashed lines represent the $\sigma(v)$ envelope, one standard
  deviation ($\pm 60$ km~s$^{-1}$, following Sandage \cite{sand86})
  about the velocity/angular--distance relation, pertaining for
  galaxies considered firmly established as members of the Local
  Group.} \label{fig:F_cost}
\end{figure}

\end{document}